\documentclass{99-Styles/MICE}
\usepackage{lineno}
\usepackage{bibentry}
\nobibliography*
\usepackage{times}
\usepackage{bm}         
\usepackage{amsmath,amssymb,bm,slashed} 
\usepackage{amssymb}    
\usepackage{graphicx}   
\usepackage{verbatim}   
\usepackage{color}      
\usepackage{subfigure}  
\usepackage{hyperref}   
\usepackage[colaction]{multicol}
\usepackage{enumerate}

\usepackage{fmtcount}   

\usepackage[square,comma,numbers,sort&compress]{natbib}

\usepackage{authblk}	

\usepackage{booktabs}
\usepackage{rotating}

\usepackage{multicol}

\begin{document}

\thispagestyle{empty}

\begin{tabular}{p{0.175\textwidth} p{0.5\textwidth} p{0.225\textwidth}}
  \hspace{-0.8cm}\leftline{\today}                                 &
  \centering{Muon Ionization Cooling Experiment}                   &
  \rightline{RAL-P-2019-003} 
\end{tabular}
\vspace{-1.0cm}\\
\rule{\textwidth}{0.43pt}

\begin{center}
  {\bf
    {\LARGE First demonstration of ionization cooling by
      the Muon Ionization Cooling Experiment} \\
  }
  \vspace{0.2cm}
  MICE collaboration \\
  \vspace{-0.0cm}
\end{center}

\makeatletter

\newcommand{\bra}[1]{\ensuremath{\langle #1 |}}   
\newcommand{\ket}[1]{\ensuremath{| #1 \rangle}}   
\newcommand{\bigbra}[1]{\ensuremath{\big\langle #1 \big|}}   
\newcommand{\bigket}[1]{\ensuremath{\big| #1 \big\rangle}}   
\newcommand{\amp}[3]{\ensuremath{\left\langle #1 \,\left|\, #2%
                     \,\right|\, #3 \right\rangle}}  
\newcommand{\sprod}[2]{\ensuremath{\left\langle #1 |%
                     #2 \right\rangle}}  
\newcommand{\ev}[1]{\ensuremath{\left\langle #1 %
                     \right\rangle}} 
\newcommand{\ds}[1]{\ensuremath{\! \frac{d^3#1}{(2\pi)^3 %
                     \sqrt{2 E_\vec{#1}}} \,}} 
\newcommand{\dst}[1]{\ensuremath{\! %
                     \frac{d^4#1}{(2\pi)^4} \,}} 
\newcommand{\tr}{\text{tr}}
\newcommand{\sgn}{\text{sgn}}
\newcommand{\diag}{\text{diag}}
\newcommand{\BR}{\text{BR}}
\newcommand{\gsim}      {\mbox{\raisebox{-0.4ex}{$\;\stackrel{>}{\scriptstyle \sim}\;$}}}
\newcommand{\lsim}      {\mbox{\raisebox{-0.4ex}{$\;\stackrel{<}{\scriptstyle \sim}\;$}}}

\renewcommand{\vec}[1]{{\mathbf{#1}}}
\renewcommand{\Re}{{\text{Re}}}
\renewcommand{\Im}{{\text{Im}}}
\newcommand{\iso}[2]{{\ensuremath{{}^{#2}}\ensuremath{\rm #1}}}
\newcommand{\eps}{{\ensuremath{\epsilon}}}
\newcommand{\draftnote}[1]{{\bf\color{red} \MakeUppercase{#1}}}
\newcommand{\panm}[1]{{\color{blue} #1}}
\providecommand{\abs}[1]{\lvert#1\rvert}
\providecommand{\norm}[1]{\lVert#1\rVert}

\def\parenbar{\mathpalette\p@renb@r}
\def\p@renb@r#1#2{\vbox{%
  \ifx#1\scriptscriptstyle \dimen@.7em\dimen@ii.2em\else
  \ifx#1\scriptstyle \dimen@.8em\dimen@ii.25em\else
  \dimen@1em\dimen@ii.4em\fi\fi \offinterlineskip
  \ialign{\hfill##\hfill\cr
    \vbox{\hrule width\dimen@ii}\cr
    \noalign{\vskip-.3ex}%
    \hbox to\dimen@{$\mathchar300\hfil\mathchar301$}\cr
    \noalign{\vskip-.3ex}%
    $#1#2$\cr}}}

%
\providecommand{\anmne}{\mbox{$\bar\nu_{\mu} \rightarrow \bar\nu_e$}} 
\providecommand{\nmne}{\mbox{$\nu_{\mu}\rightarrow\nu_e$}} 
\providecommand{\anm}{\mbox{$\bar\nu_\mu$}} 
\providecommand{\nm}{\mbox{$\nu_\mu$}}
\providecommand{\nue}{\mbox{$\nu_e$}} 
\providecommand{\ane}{\mbox{$\bar\nu_e$}} 
\providecommand{\enu}{\mbox{$E_\nu$}}
\providecommand{\piz}{\mbox{$\pi^0 $}}
\providecommand{\pip}{\mbox{$\pi^+$}} 
\providecommand{\pim}{\mbox{$\pi^-$}}

\parindent 10pt
\pagestyle{plain}
\pagenumbering{arabic}                   
\setcounter{page}{1}

\noindent\textbf{%
High-brightness muon beams of energy comparable to those produced by state-of-the-art electron, proton and ion accelerators have yet to be realised. Such beams have the potential to carry the search for new phenomena in lepton-antilepton collisions to extremely high energy and also to provide uniquely well-characterised neutrino beams. A muon beam may be created through the decay of pions produced in the interaction of a proton beam with a target. To produce a high-brightness beam from such a source requires that the phase space volume occupied by the muons be reduced (cooled). Ionization cooling is the novel technique by which it is proposed to cool the beam. The Muon Ionization Cooling Experiment collaboration has constructed a section of an ionization cooling cell and used it to provide the first demonstration of ionization cooling. We present these ground-breaking measurements.
}

\begin{multicols}{2}
  
\noindent
Fundamental insights into the structure of matter and the nature of
its elementary constituents have been obtained using beams of charged
particles.
The use of time-varying electromagnetic fields to produce sustained
acceleration was pioneered in the
1930s~\cite{Lawrence:1931cb,Lewis:1933zz,Lawrence1934,Lawrence:1936uw,Wideroe:1947egh,Wideroe:1953ufn}.
Since then, high-energy and high-brightness particle accelerators
have delivered electron, proton, and
ion beams for applications that range from the search for new
phenomena in the interactions of quarks and leptons, to the study of
nuclear physics, materials science, and biology.

Muon beams are created using a proton beam striking a target to produce a secondary
beam comprising many particle species including pions, kaons and muons. The pions 
and kaons decay to produce additional muons that are captured 
by electromagnetic beamline elements to produce a tertiary muon beam. Capture and
acceleration must be realised on a time scale compatible with the
$2.2\,\mu$s muon lifetime at rest.
The energy of the muon beam is limited by the
energy of the primary proton beam and the intensity is
limited by the efficiency with which muons are accepted into the
transport channel. 
High-brightness muon beams have not yet been produced at energies
comparable to state-of-the-art electron and proton beams.

Accelerated high-brightness muon beams have been proposed as a
source of neutrinos at a neutrino factory and to deliver
multi-TeV lepton-antilepton collisions at a muon
collider~\cite{Neuffer:1994bt,Geer:1997iz,Apollonio:2002en,PhysRevSTAB.6.081001,Palmer:2014nza,Boscolo:2018tlu,Neuffer2018}.
Muons have properties that make them ideal candidates for the delivery of high energy collisions.
The muon is a fundamental particle with mass 207 times that of the electron, making
collisions possible between beams of muons and anti-muons at energies far in excess of
those that can be achieved in an electron-positron collider such as the proposed International Linear Collider~\cite{Behnke:2013xla}, the Compact Linear Collider~\cite{Charles:2018vfv,Roloff:2018dqu,Aicheler:2019dhf} or the electron-positron option of the Future Circular Collider~\cite{Abada:2019zxq}. 
The energy available in collisions between the constituent gluons and quarks in proton-proton 
collisions is significantly less than the
proton-beam energy because the colliding quarks and gluons each carry only a fraction
of the proton's momentum. This makes muon colliders attractive to take the study of
particle physics beyond the reach of facilities such as the Large Hadron Collider~\cite{MYERS:2013hra}. 

Most of the proposals for accelerated muon beams exploit the proton-driven muon beam
production scheme outlined above.
In these proposals the tertiary muon beam has its brightness increased through beam cooling
before it is accelerated and stored.
Four cooling techniques are in use at particle accelerators:
synchrotron radiation cooling \cite{2012acph.book.Lee}; laser
cooling~\cite{PhysRevLett.64.2901,PhysRevLett.67.1238,doi:10.1063/1.329218};
stochastic cooling~\cite{Mohl:1980jb,Marriner:2003mn}; and electron
cooling~\cite{1063-7869-43-5-R01}. In each case the time taken to cool the beam is
long compared to the muon lifetime.
Frictional cooling of muons, in which muons are electrostatically accelerated 
through an energy-absorbing medium at energies significantly below an MeV, 
has been demonstrated but only with low efficiency~\cite{1999HyInt.119..305M, 2005NIMPA.546..356A, 2006PhRvL..97s4801T,PhysRevLett.112.224801}.

The novel technique demonstrated in this paper, 
ionization cooling~\cite{Skrinsky:1981zz,Neuffer:1983jr}, is expected to 
occur when a suitably prepared beam passes through 
an appropriate material (the absorber) and loses momentum through ionization.
Radio-frequency cavities restore momentum along the beam 
direction only.
Passing the muon beam through a repeating lattice of material and accelerators 
causes the ionization cooling effect to build up
in a time much shorter than the muon 
lifetime~\cite{Rogers:2013kna,Stratakis:2014nna,Neuffer:2016gtw}.
Acceleration of a muon beam in a radio-frequency accelerator has recently
been demonstrated ~\cite{Bae:2018atj} 
and reduced beam heating, damped by the ionization cooling effect, has
been observed~\cite{Mori:2009zz}. However, ionization cooling has never
previously been demonstrated. Such a confirmation 
is important for the development of future muon accelerators.
The international Muon Ionization Cooling Experiment
(MICE)~\cite{MICE-WWW} was designed to demonstrate
transverse ionization cooling, the first observation of which is
presented here.

  The brightness of a particle beam can be characterised by
the number of particles in the beam and the volume occupied by the
beam in position-momentum phase space. 
The phase space considered in this paper is the position and momentum
transverse to the direction of travel of the beam: 
$\vec{u} = (x, p_x, y, p_y)$, where $x$ and $y$ are coordinates perpendicular
to the beam line, and $p_x$ and $p_y$ are the corresponding components 
of momentum. The $z$-axis is the nominal beam axis.

The phase space volume occupied by the beam
and the phase space density of the beam are conserved quantities in
a conventional accelerator without cooling.  
The normalised root-mean-square (RMS) emittance is often used as an indicator of the
phase space volume occupied by the beam and is given by 
\cite{Penn:2000mt}

\begin{equation}
  \varepsilon_{\perp} = \frac{\sqrt[4]{|\mathbf{V}|}}{m_\mu} \, ,
\end{equation}

\noindent 
where $m_\mu$ is the muon mass and $|\mathbf{V}|$ is the determinant of the covariance
matrix of the beam in transverse phase space. The covariance matrix has elements $v_{ij} = \left<u_i u_j\right> - 
\left<u_i\right> \left<u_j\right>$. The distribution of individual particle amplitudes 
also describes the volume of the beam in phase space. 
The amplitude is defined by~\cite{holzer2004}

\begin{equation}
  \label{eq:amplitude}
  A_\perp = \varepsilon_{\perp} R^2(\vec{u}, \left<\vec{u}\right>)\, ,
\end{equation}

\noindent 
where $R^2(\vec{u}, \vec{v})$ is the square of the distance 
between two points, $\vec{u}$ and $\vec{v}$, 
in the phase space, normalised to the covariance matrix:

\begin{equation}
R^2(\vec{u}, \vec{v}) = (\vec{u} - \vec{v})^T \, \mathbf{V^{-1}} \, (\vec{u} - \vec{v}).
\end{equation}

\noindent 
The normalised RMS emittance is proportional to the mean of the
particle amplitude distribution.
In the approximation that particles travel near to the beam axis, and
in the absence of cooling, the particle amplitudes and the normalised
RMS emittance are conserved quantities. If the beam is well described 
by a multivariate Gaussian distribution then $R^2$ is distributed 
according to a $\chi^2$ distribution with four degrees of freedom so the 
amplitudes are distributed according to

\begin{equation}
  f(A_\perp) = \frac{A_\perp}{4\varepsilon_\perp^2} \; \mathrm{exp}\!\left(\frac{-A_\perp}{2\varepsilon_\perp}\right).
\end{equation}

\noindent 
The rate of change of the normalised transverse emittance as the beam
passes through an absorber is given approximately
by~\cite{Neuffer:1983jr,Penn:2000mt,Rogers08beamdynamics}

\begin{equation}
  \frac{d\varepsilon_{\perp}}{dz}\backsimeq
  -\frac{\varepsilon_{\perp}}{\beta^2E_{\mu}}\left|
  \frac{dE_{\mu}}{dz} \right| +
  \frac{\beta_{\perp}(13.6\,\text{MeV/}c)^2}{2\beta^3E_{\mu}m_{\mu}X_0} \, ,
  \label{eq:cooling}
\end{equation}

\noindent 
where $\beta c$ is the muon velocity, $E_{\mu}$ the energy,
$\left| \frac{dE_{\mu}}{dz} \right|$ the mean energy loss per unit 
path length, $X_0$ the radiation length of the absorber and 
$\beta_{\perp}$ the transverse betatron function at the
absorber~\cite{Penn:2000mt}. 
The first term of this equation describes `cooling' by ionization
energy loss and the second describes `heating' by multiple Coulomb
scattering. 
Equation \ref{eq:cooling} implies that there is an equilibrium emittance
for which the emittance change is zero.

If the beam is well described by a multivariate gaussian
distribution both before and after cooling then the 
downstream and upstream amplitude distributions  $f^d(A_\perp)$ and $f^u(A_\perp)$
are related to the downstream and upstream emittances $\varepsilon_\perp^d$ and $\varepsilon_\perp^u$ by

\begin{equation}
\frac{f^d(A_\perp)}{f^u(A_\perp)} = 
\left(\frac{\varepsilon_{\perp}^{u}}{\varepsilon_{\perp}^{d}}\right)^2
\exp\left[-\frac{A_\perp}{2}\left(\frac{1}{\varepsilon_\perp^d}-\frac{1}{\varepsilon_\perp^u}\right)\right]\,.
\end{equation}

Many particles in the experiment described in this paper do not travel
near to the beam axis. These particles experience effects from optical 
aberrations, as well as geometrical effects such as scraping, in which
high amplitude particles outside the experiment's aperture are removed
from the beam. Scraping reduces the emittance of the ensemble, and 
selectively removes those particles that scatter more than the rest of the 
ensemble. Optical aberrations and scraping introduce a bias in the change in
RMS emittance that occurs due to ionization cooling. 
In this paper the distribution of amplitudes is studied. 
In order to expose the behaviour in the beam core, independently of
aberrations affecting the beam tail, $\mathbf{V}$ and
$\varepsilon_\perp$ are recalculated for each amplitude bin, including 
particles that are in lower amplitude bins and excluding particles that are in 
higher amplitude bins. This results in a distribution that, in the core of the beam, is
independent of scraping effects and aberrations.

Change in phase space density provides a direct measurement of the cooling 
effect. The $k$-Nearest Neighbour ($k$NN) algorithm provides a robust 
non-parametric estimator of the phase space density of the muon
ensemble~\cite{Mack19791, drielsma:2018}.
The separation of pairs of muons is characterised by the normalised squared
distance, $R_{ij}^2(\vec{u}_i, \vec{u}_j)$, between muons with position
$\vec{u_i}$ and $\vec{u_j}$. The density, $\rho_i$, associated with the $i^{\rm{th}}$ particle is estimated by

\begin{equation}
  \rho_i(\vec{u_i}) = \frac{k}{n|\mathbf{V}|^\frac{1}{2}} 
  \frac{1}{v_{ik}} = \frac{2k}{n\pi^2|\mathbf{V}|^\frac{1}{2}} 
  \frac{1}{R_{ik}^4} \, ,
\end{equation}

\noindent 
where $v_{ik}$ is the volume of the hypersphere, centred on
$\vec{u}_i$, that intersects the particle having the $k^{th}$
smallest $R_{ij}$ and $n$ is the number of particles in the ensemble.  
An optimal value for $k$ has been used, $k = n^{4/(4+d)} = \sqrt{n}$, 
with phase space dimension $d=4$ \cite{drielsma:2018}.

  \begin{figure*}[!tbh]
  \hspace*{1.25cm}\includegraphics*[width=0.97\textwidth]{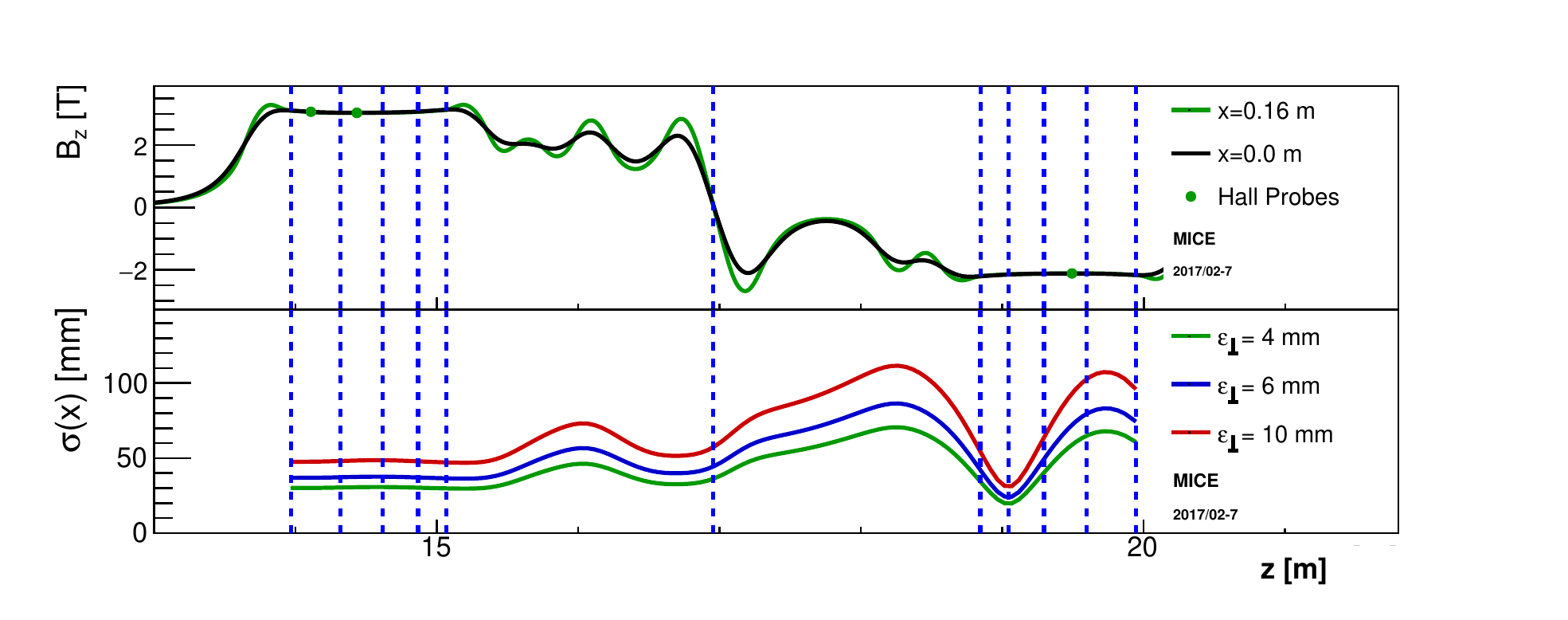}
  \includegraphics*[width=1.0\textwidth]{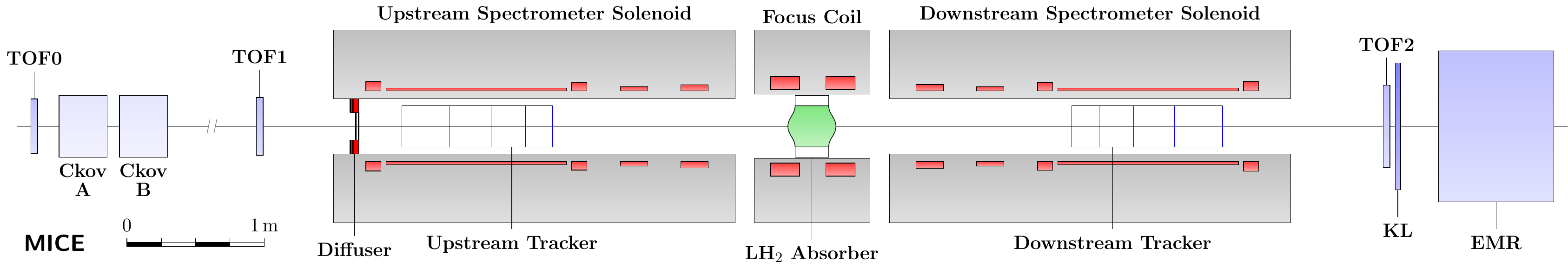}
  \caption{
    The MICE apparatus along with the calculated
    magnetic field, $B_z$\,[T], and nominal horizontal width of the beam, $\sigma(x)$\,[mm]. 
    The modelled field is shown on the beam axis and 160 mm from the axis in the horizontal plane. 
    The readings of Hall probes, situated 160 mm from the beam axis,
    are also shown. 
    Dashed lines indicate the position of the tracker stations and
    absorber. 
    The nominal RMS beam width is calculated assuming a nominal input
    beam using linear beam transport equations. 
    Acronyms used in the schematic are described in the text. 
  }
  \label{fig:Step4}
\end{figure*}

The MICE collaboration has built a tightly focusing solenoid lattice, absorbers and 
instrumentation to demonstrate ionization cooling of muons.
A schematic of the apparatus is shown in figure~\ref{fig:Step4}.

A transfer line \cite{Bogomilov:2012sr, Adams:2013lba, Adams:2015wxp}
brought a beam, composed mostly of muons, from a target~\cite{Booth_2016} in the ISIS
synchrotron~\cite{THOMASON201961} to the cooling apparatus. 
The muons had a nominal momentum of 140~MeV/c.
A variable thickness brass and tungsten diffuser allowed the incident beam 
emittance to be varied between 4 and 10 mm.

The tight focussing (low beta function) and large acceptance required by the cooling section was
achieved using twelve superconducting solenoids. The solenoids were contained in three warm-bore
modules cooled by closed cycle cryocoolers. The upstream and downstream modules (the `spectrometer solenoids') were identical, each containing three coils to provide a uniform field region of up to
4T within the 400\,mm diameter warm bore for momentum measurement, and two `matching' coils to match
the beam to the central pair of closely spaced `focus' coils which focussed the beam onto the
absorber. The focus coils were a pair of split-field coils designed for peak on-axis fields of up to
3.5\,T contained within one module with a 500\,mm diameter warm bore which contained the
absorbers. For the data reported here the focus coils were operated in 'flip' mode with a field
reversal at the centre. Because the magnetic lattice was tightly coupled the cold mass suspension 
systems of the modules were designed to withstand the longitudinal cold-to-warm forces of several 
hundred kN which could arise during an unbalanced quench of the system. At maximum field the 
inter-coil force on the focus coil cold mass was of the order of 2\,MN. The total energy stored in 
the magnetic system was of the order of 5\,MJ and the system was protected by both active and 
passive quench protection systems. The normal charging and discharging time of the solenoids was 
several hours. The entire magnetic channel was partially enclosed by a 
150\,mm thick soft-iron return yoke for external magnetic shielding. The magnetic fields in the 
tracking volumes were monitored during operation with calibrated Hall probes.

One of the matching coils in the downstream spectrometer solenoid was not operable due to a failure of a superconducting lead. While this necessitated a compromise in the lattice optics and acceptance, the flexibility of the magnetic lattice was exploited to ensure a clear cooling measurement.

The amplitude acceptance of approximately 30\,mm, above which particles scrape, 
was large compared to a typical accelerator. Even so 
significant scraping was expected and observed for the highest emittance beams. Ionization
cooling cells with even larger acceptances, producing less scraping, have
been designed~\cite{Rogers:2013kna,Stratakis:2014nna,Neuffer:2016gtw}.
The magnetic lattice of MICE was tuned so that 
the beam had a focus near
to the absorber resulting in a small beam width, shown in figure~\ref{fig:Step4},
and large angular divergence.
The tight focussing, corresponding to a region of small $\beta_\perp$, yielded an optimal
cooling performance, as implied by equation \ref{eq:cooling}.

Materials with low atomic number
such as lithium and hydrogen have a long radiation length 
relative to the rate of energy loss and consequently low equilibrium
emittance, making them ideal absorber materials.
Therefore the cooling due to both liquid hydrogen and lithium hydride absorbers was studied.

The liquid hydrogen was contained within a 22\,l vessel~\cite{Bayliss:2018rkd}
in the warm bore of the focus coil. Hydrogen was liquefied by a cryocooler and 
piped through the focus coil module into the absorber body.
When filled, the absorber presented $349.6 \pm 0.2$\,mm of liquid
hydrogen along the beam axis with a density of
$0.07053 \pm 0.00008$\,g/cm$^3$. 
The liquid hydrogen was contained by a pair of
aluminium windows covered by multi-layer insulation. 
A second pair of windows provided secondary containment to protect
against the possibility
of failure of the primary containment windows. The total thickness
of all four windows on the beam axis was $0.79 \pm 0.01$\,mm.

The lithium hydride absorber was a $65.37 \pm 0.02$\,mm thick disk
with a density of $0.6957 \pm 0.0006$\,g/cm$^3$. The isotopic
composition of the lithium used to produce the absorber 
was 95 $\%$ $^6$Li and 5 $\%$ $^7$Li.
The cylinder had a thin coating of parylene to prevent ingress of
water or oxygen. 
Configurations with no absorber installed at all and with the empty
liquid hydrogen containment vessel were also studied. 

Detectors placed upstream and downstream of the apparatus measured the
momentum, position, and species of each particle entering and leaving
the cooling channel so that the full four-dimensional phase space,
including the angular momentum introduced by the solenoids, could be
reconstructed.  
Particles were recorded by the apparatus one at a time, which enabled 
high-precision instrumentation to be used and particles other than muons
to be excluded from the analysis. 
Each ensemble of muons was accumulated over a number of
hours of operation of the experiment. 
This is acceptable as collective effects are not expected at a neutrino
factory and in a muon collider collective effects become significant only
at very low longitudinal emittance~\cite{PhysRevSTAB.18.044201}.
Data-taking for each absorber was
separated by a period of weeks due to operational practicalities.
The phase space distribution of the resulting ensemble was
reconstructed using the upstream and downstream detectors. 
Emittance reconstruction in the upstream detector system is
described in \cite{Adams2019}. 

Upstream of the cooling apparatus, two time-of-flight detectors
(TOFs)~\cite{Bertoni:2010by,Bertoni:2010bb} measured particle velocity.  
A complementary velocity measurement was made upstream by threshold
Cherenkov counters Ckov\,A and Ckov\,B~\cite{Cremaldi:2009}.
Scintillating fibre trackers, positioned in the uniform-field region
of each of the two spectrometer solenoids, measured
particle position and momentum upstream and downstream of the
absorber~\cite{Ellis:2010bb,Dobbs:2016ejn,Adams:2013lba}.
Downstream, an additional TOF detector, a mixed lead and scintillator pre-shower
detector (KL), and a totally active scintillator calorimeter, the
Electron Muon Ranger (EMR)~\cite{Adams:2015eva,Asfandiyarov:2016erh}
identified electrons produced in muon decay and allowed cross-validation
of the measurements made by the upstream detectors and the trackers.

Each tracker consisted of five planar scintillating fibre
stations. 
Each station comprised three views, each view composed of
scintillating fibres laid at an angle of 120$^\circ$ with respect to
the other views.
Each view was made of two layers of 350\,$\mathrm{\mu}$m diameter scintillating
fibres. 
Groups of seven scintillating fibres were read out together by
cryogenic Visible Light Photon Counters~\cite{VLPC,VLPC1}.
The position of a particle crossing the tracker was inferred from the coincidence
of signals from the fibres and momentum was inferred by fitting a helical
trajectory to the positions with appropriate consideration for energy loss
and scattering in the fibres.

Each TOF was constructed from two orthogonal planes of scintillator slabs.
Photomultiplier tubes at each end of every TOF slab were used to
determine the time at which a muon passed through the apparatus with a
$60$\,ps resolution~\cite{Bertoni:2010by}. 
The momentum resolution of particles for which the radius of the helix in the
tracker was small was improved by combining the TOF
measurement of velocity with the measurement of momentum in the
tracker.

A detailed Monte Carlo simulation of the experiment was performed to
study the resolution and efficiency of the instrumentation and to
determine the expected performance of the 
cooling apparatus~\cite{Agostinelli:2002hh,Alli06,Asfandiyarov:2018yds}.
The simulation was found to give a good description of the data \cite{Adams2019}.

\begin{figure*}[!tbh]
  \includegraphics[width=\textwidth]{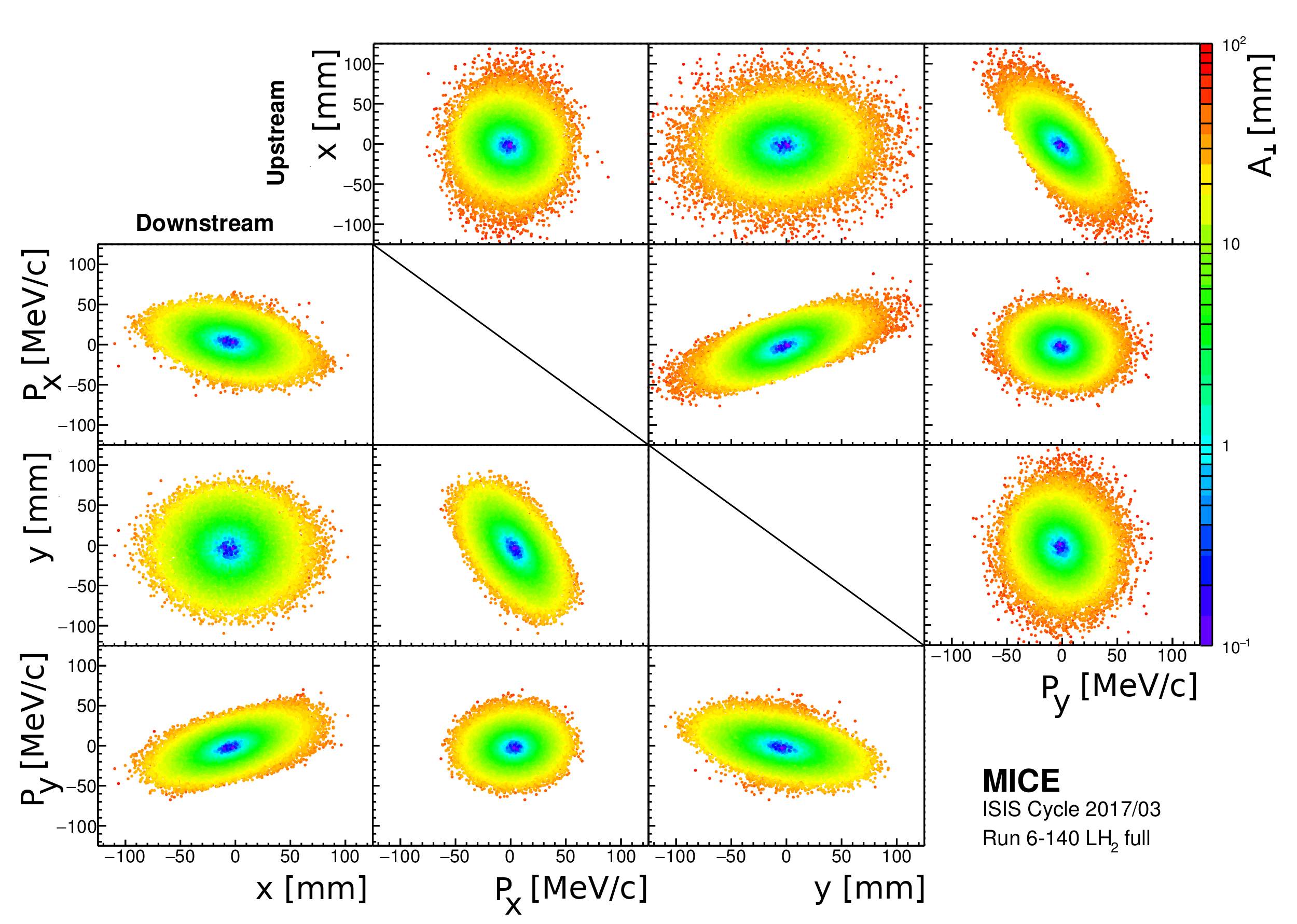}
  \caption{
    Distribution of the beam in phase space for the 6-140 Full LH$_2$
    setting: (above the diagonal) measured in the upstream tracker and (below the
    diagonal) measured in the downstream tracker. 
    Measured particles' positions are shown, coloured according to the
    amplitude of the particle. 
  }
  \label{fig:phase_space}
\end{figure*}

The data presented here were taken using beams with a nominal momentum of 140\,MeV/c and
with a nominal normalised RMS emittance in the upstream tracking volume of 4\,mm,
6\,mm and 10\,mm. These settings are denoted `4-140', `6-140' and `10-140' respectively.
Beams with a higher emittance have correspondingly higher amplitude
and occupy a larger region in phase space.
For each beam setting, two samples were considered for the analysis.
The `upstream sample' contained particles identified as muons using
the upstream TOF detectors and tracker, for which the muon trajectory reconstructed in the
upstream tracker was fully contained in the fiducial volume and for
which the reconstructed momentum fell within the range 135\,MeV/c to
145\,MeV/c, which was significantly larger than the 2\,MeV/c momentum 
resolution of the tracker.
The `downstream sample' was that subset of the upstream sample for
which the reconstructed muons were fully contained in the fiducial
volume of the downstream tracker. The samples each had between 30,000 and
170,000 events.
The distributions in phase space of the particles in the two samples are
shown in figure~\ref{fig:phase_space}. 
The strong correlations between $y$ and $p_x$ and between $x$ and $p_{y}$
are due to the angular momentum introduced by the
solenoidal field. The shorter tail along the semi-minor axis
than the semi-major axis in these projections arises from
scraping in the diffuser.

  \begin{figure*}[!tbh]
  \centering
  \hspace*{-1.0cm}
  \includegraphics[width=1.1\textwidth]{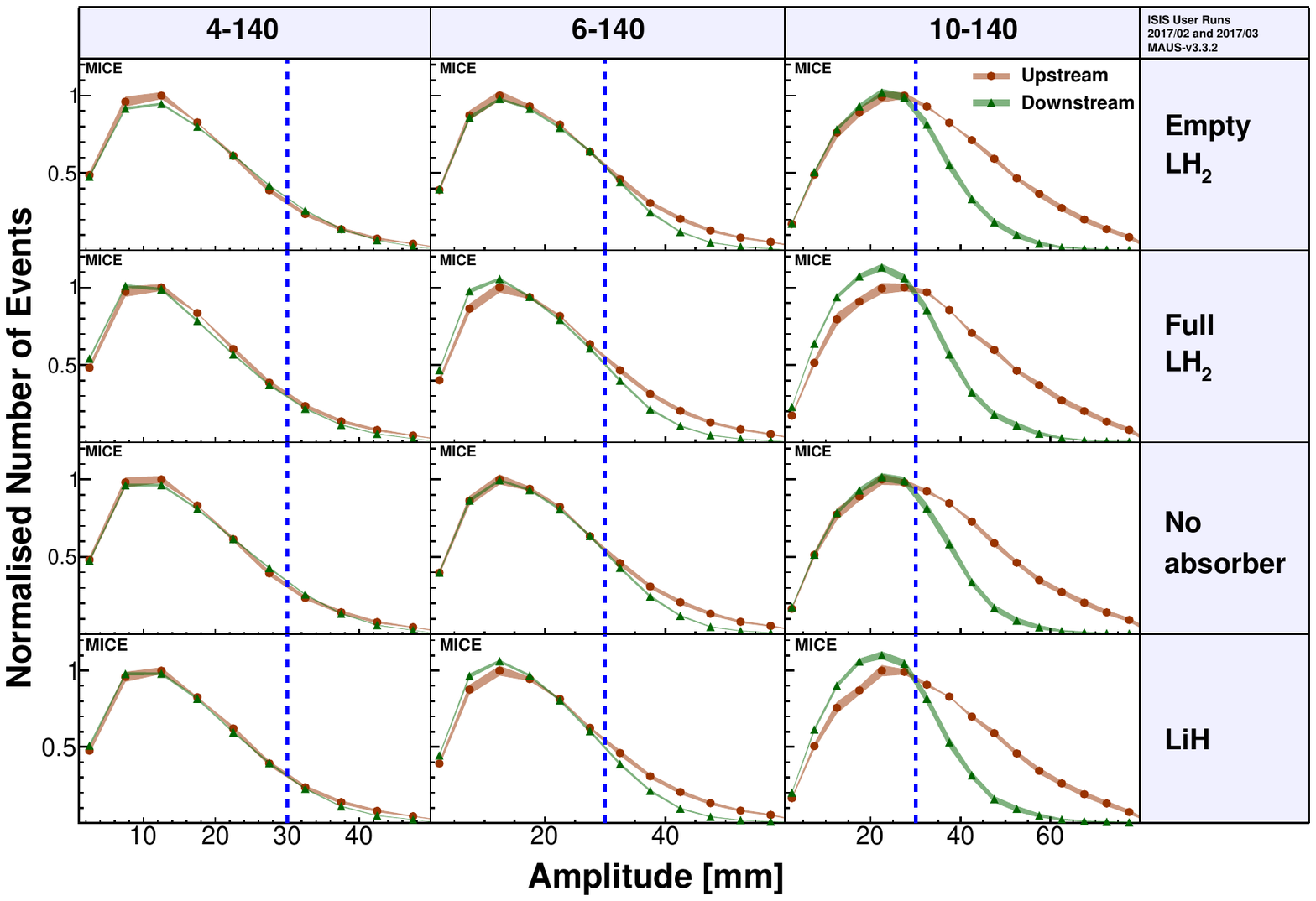}
  \caption{
    The distributions of measured muon amplitudes. 
    The upstream distributions are shown by orange circles while the
    downstream distributions are shown by green triangles. 
    Both upstream and downstream distributions are normalised to the
    bin in the upstream distribution with the most entries (see text). 
    Coloured bands show the uncertainty, which is dominated by systematic uncertainties. 
    Vertical lines indicate the approximate channel acceptance above which
    scraping occurs.
  }
  \label{fig:amplitude_pdf_reco}
\end{figure*}

The distribution of amplitudes in the upstream and downstream samples
for each of the 4-140, 6-140, and 10-140 data sets is shown in
figure~\ref{fig:amplitude_pdf_reco}.
The nominal acceptance of the magnetic channel is also indicated.
A correction has been made to account for the migration of events
between amplitude bins that arises due to the detector resolution and
to account for inefficiency in the downstream detector system. The 
correction is described in the Methods section.
Distributions are shown for the case where there was no absorber (`No
absorber'), where the liquid hydrogen vessel was empty (`Empty LH$_{2}$'),
where the liquid hydrogen vessel was filled (`Full LH$_{2}$'), and where the
lithium hydride absorber was present (`LiH').
The distributions were normalised to allow a 
comparison of the shape of the distribution between different absorbers.
Each pair of upstream and downstream amplitude distributions is scaled
by $1/N^{u}_{max}$, where $N^{u}_{max}$ is the number of 
events in the most populated bin in the upstream sample.

The behaviour of the beam at low amplitude is the key result of this paper.
For the `No absorber' and the `Empty
LH$_{2}$' configurations, the number of events with low amplitude in the 
downstream sample is similar to that observed in the upstream sample.
For the 6-140 and 10-140 configurations for both the `Full LH$_{2}$' and 
the `LiH' samples, the number of events with low amplitude is significantly larger 
in the downstream sample than in the upstream sample. This indicates
an increase in the number of particles in the beam core when an absorber is
installed, which is expected if ionization cooling occurs.
This effect can only occur because energy loss due to
ionization is a non-conservative process.

A reduction in the number of muons at high amplitude is also
observed, especially for the 10-140 setting.
While some of this effect arises due to migration of muons into the
beam core, a significant number of high amplitude particles migrated away from the
beam acceptance due to optical mismatch and were scraped on
apertures.

A $\chi^2$ test was performed to determine the confidence with 
which the null hypothesis that, for the same input beam setting, the amplitude distribution in the downstream 
samples of the `Full LH$_{2}$' and `Empty LH$_{2}$' configurations are compatible, and the 
amplitude distribution in the downstream samples of the `LiH' and `No absorber' 
configurations are compatible. The test was performed on the uncorrected distributions
assuming statistical uncertainties only. Systematic effects are the same for the
pairs of distributions tested and cancel. The probability of observing the effect seen in 
the data, assuming this null hypothesis is correct, is significantly less
than $10^{-5}$ for all beam settings and all pairs of `Full LH$_{2}$' and `Empty LH$_{2}$' and all pairs of `LiH' and `No absorber', therefore the null hypothesis was rejected.
\begin{figure*}[!tbh]
  \begin{center}
    \hspace*{-1.0cm}
    \includegraphics[width=1.1\textwidth]{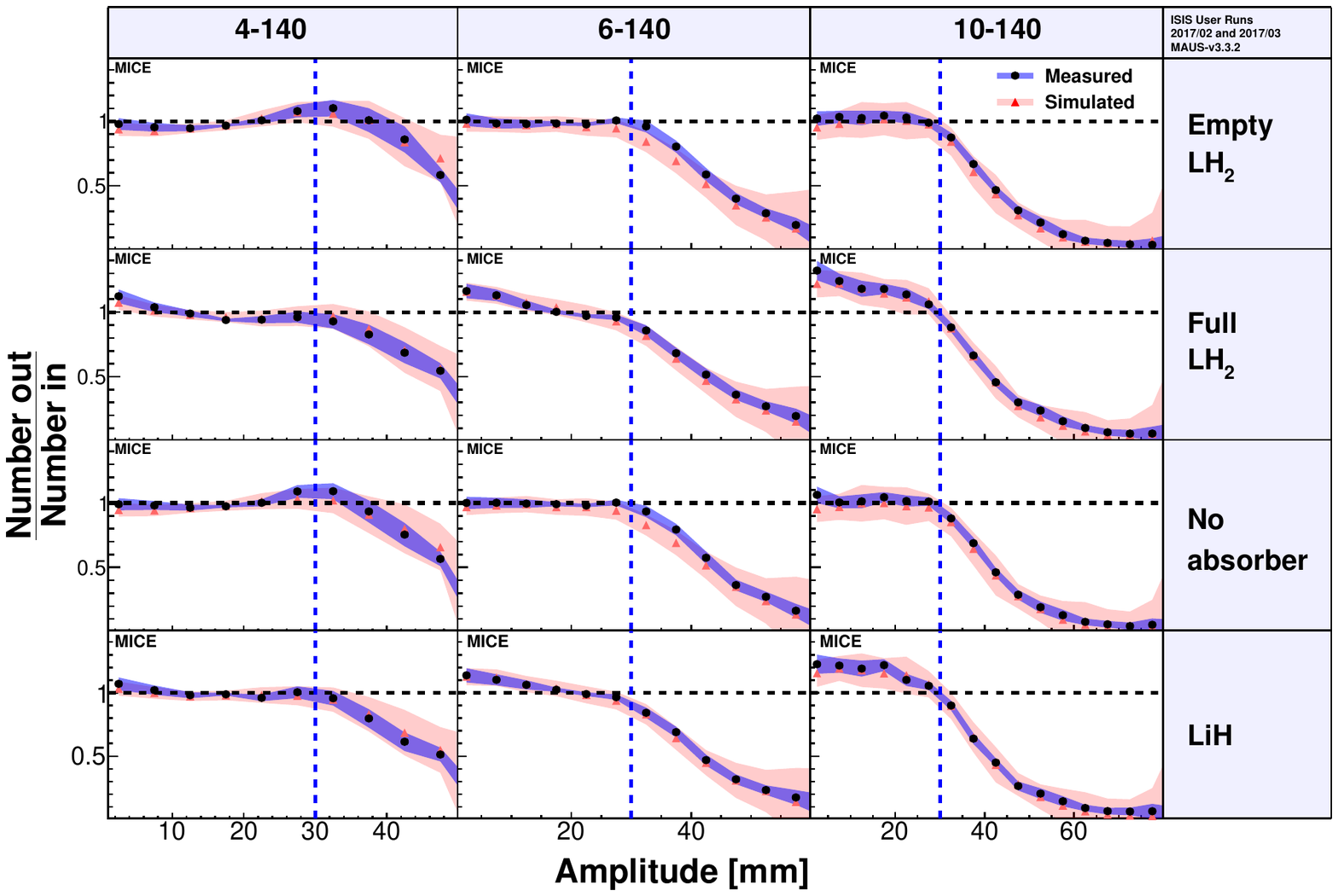}
  \end{center}
  \caption{
    Downstream to upstream ratio of number of events. 
    A ratio greater than unity in the beam core is evidence for ionization
    cooling and is evident for 6-140 and 10-140 beams with both the full LH$_{2}$ absorber and the LiH absorber. 
    The effect predicted from simulation is shown in red, while that
    measured is shown in black. 
    Uncertainty is shown by a blue fill for data and a pink fill for 
    simulation  and is dominated by systematic uncertainty.
    Vertical lines indicate the channel acceptance above which
    scraping occurs.
  }
  \label{fig:pdf_ratio}
\end{figure*}

The fractional increase in the number of particles with low amplitude is 
most pronounced for the 10-140 beams.
High amplitude beams have high $\varepsilon_\perp$ and a larger
transverse momentum relative to the stochastic increase in transverse momentum 
due to scattering, so undergo
more cooling, as predicted by equation \ref{eq:cooling}. 
For the magnet settings and beams studied here the equilibrium emittance of the
experiment is close to 4\,mm.
As a result only modest cooling is observed for the 4-140 setting in 
both the `Full LH$_{2}$' and the `LiH' configuration.

The ratio of the downstream to the upstream amplitude distribution is
shown in figure~\ref{fig:pdf_ratio}. 
In the `No absorber' and `Empty absorber' configurations, the ratio is
consistent with 1 for amplitudes less than 30\,mm, confirming the
conservation of amplitude in this region irrespective of the incident
beam.
Above 30\,mm the ratio drops below unity, indicating that there are fewer
muons downstream than upstream due to the beam scraping on apertures.
The presence of the absorber windows does not strongly affect the
amplitude distribution. 
The liquid hydrogen absorber windows were designed to be as thin as 
possible so that when installed, scattering in the windows would not 
cause significant heating.
For the 6-140 and 10-140 data sets, the addition of liquid hydrogen or 
lithium hydride absorber material causes the ratio to rise above unity
for low amplitude particles, corresponding to the beam core.
This indicates an increase in the number of particles in the beam
core and demonstrates ionization cooling.

\begin{figure*}[!tbh]
  \begin{center}
    \hspace*{-1.0cm}
    \includegraphics[width=1.1\textwidth]{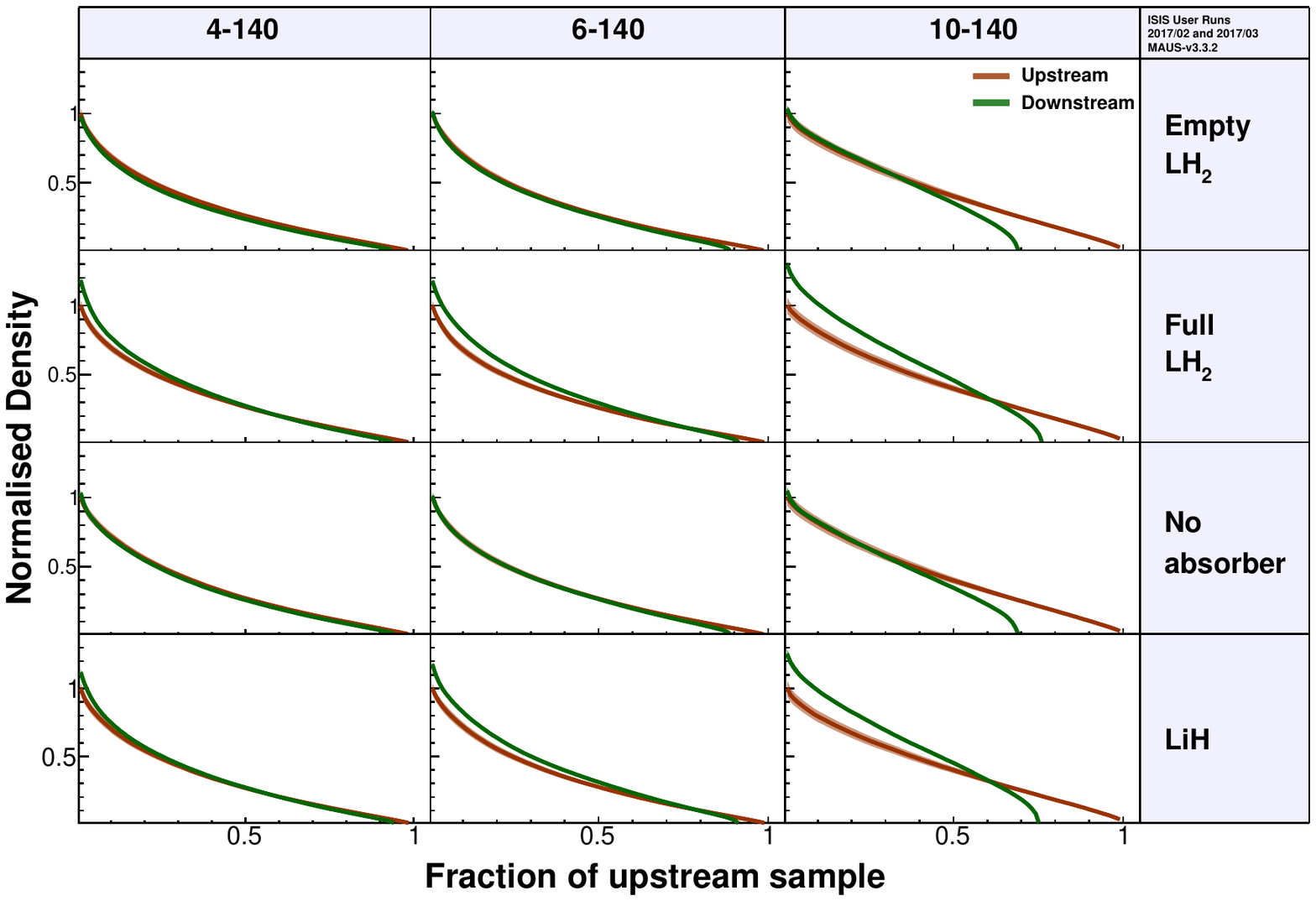}
  \end{center}
  \caption{
    The upstream and downstream normalised beam density quantiles, indicated 
    by orange and green lines respectively, as a function 
    of the fraction of the upstream sample. For each configuration, the density is normalised to 
    the highest density region in the upstream sample. Uncertainty is indicated by
    the thickness of the coloured bands and is dominated by systematic uncertainty.
  }
  \label{fig:density}
\end{figure*}

The density in phase space is an invariant of a symplectic system, therefore an increase
in phase space density is also an unequivocal demonstration of cooling. 
Figure~\ref{fig:density} shows the normalised density of the beam $\rho_i(\vec{u_i})/\rho_0$ 
as a function of $\alpha$, the fraction of the upstream sample that has a 
density greater than or equal to $\rho_i$.
To enable comparison between different beam configurations, the densities for each 
configuration have 
been normalised to the peak density in the upstream tracker, $\rho_0$. 
To enable comparison between upstream and downstream distributions, the fraction of the sample is 
always relative to the total number of events in the upstream sample. The transmission is
the fraction of the beam where the density in the downstream tracker reaches zero.
For the `No absorber' and `Empty LH$_{2}$' cases the density downstream in the highest density 
regions is indistinguishable from the density upstream. 
A small amount of scraping is observed for the 4-140 and 6-140 beams. More significant scraping is 
observed for the 10-140 beam. In all cases, for `Full LH$_{2}$' and `LiH', the phase space 
density increases. 
The increase is more significant for higher emittance beams.
These observations demonstrate the ionization cooling of the beam when
an absorber is installed. In the presence of an absorber, beams with larger nominal emittance show
a greater increase in density than those with a lower nominal
emittance, which is consistent with equation \ref{eq:cooling}.

Ionization cooling has been unequivocally demonstrated for the first time. 
The MICE collaboration has built and operated a section of solenoidal cooling channel
and demonstrated the ionization cooling of muons using both
liquid hydrogen and lithium hydride absorbers. The effect has been observed both 
from observation of an increase in the number of particles having a small amplitude
(figures \ref{fig:amplitude_pdf_reco} and  \ref{fig:pdf_ratio}) and an increase 
in the phase space density of 
the beam (figure \ref{fig:density}). The results agree well with simulation (figure \ref{fig:pdf_ratio}).
The ground-breaking demonstration of ionization cooling presented here is a significant advance 
in the development of high-brightness muon beams. 
The seminal results presented in this paper encourage further development
of high-brightness muon beams as a tool for the investigation of the 
fundamental properties of matter.

  \section*{Acknowledgements}

The work described here was made possible by grants from the 
Science and Technology Facilities Council (UK), the Department of
Energy and the National Science Foundation (USA), the Instituto Nazionale
di Fisica Nucleare (Italy), the European Community under the European Commission
Framework Programme 7 (AIDA project, grant agreement no. 262025, TIARA
project, grant agreement no. 261905, and EuCARD), the Japan Society
for the Promotion of Science, the National Research Foundation of 
Korea (No. NRF-2016R1A5A1013277), the Ministry of Education, Science 
and Technological Development of the Republic of Serbia, the Institute of High
Energy Physics/Chinese Academy of
Sciences fund for collaboration between the People's Republic of China and the USA
and the Swiss National Science Foundation, in the framework of the SCOPES programme. 
We gratefully acknowledge all sources of support.
We are grateful for the support given to us by the staff of the STFC
Rutherford Appleton and Daresbury Laboratories.
We acknowledge the use of Grid computing resources deployed and
operated by GridPP in the UK, \url{http://www.gridpp.ac.uk/}.

\section*{Data Availability}
The unprocessed and reconstructed data that support the findings of this study are publicly available on the GridPP computing Grid via the data DOIs:
\begin{itemize}
\item The MICE unprocessed data: \url{https://doi.org/doi:10.17633/rd.brunel.3179644}.
\item The MICE reconstructed data: \url{https://doi.org/doi:10.17633/rd.brunel.5955850}.
\end{itemize}
Publications using the MICE data must contain the following statement:
\emph{We gratefully acknowledge the MICE collaboration for allowing us access to their data. Third-party results are not endorsed by the MICE collaboration.}

\section*{Software Availability}
The MAUS software \cite{Asfandiyarov:2018yds} that was used for reconstructing and analysing the MICE data is available at \url{https://doi.org/doi:10.17633/rd.brunel.8337542}.

\section*{Authorship}
All authors contributed significantly to the design or construction of the apparatus or to the data-taking or analysis described here.
\end{multicols}

\bibliographystyle{99-Styles/utphys}
\bibliography{Ampl-evolve}

\providecommand{\href}[2]{#2}\begingroup\raggedright\begin{thebibliography}{10}

\bibitem{Lawrence:1931cb}
E.~O. Lawrence and M.~S. Livingston, ``{The production of high speed protons
  without the use of high voltages},''
  \href{http://dx.doi.org/10.1103/PhysRev.38.834}{{\em Phys. Rev.} {\bfseries
  38} (1931) 834}.

\bibitem{Lewis:1933zz}
G.~N. Lewis, M.~S. Livingston, and E.~O. Lawrence, ``{The Emission of
  Alpha-Particles from Various Targets Bombarded by Deutons of High Speed},''
  \href{http://dx.doi.org/10.1103/PhysRev.44.55}{{\em Phys. Rev.} {\bfseries
  44} (1933) 55--56}.

\bibitem{Lawrence1934}
E.~O. Lawrence, ``Method and apparatus for the acceleration of ions,'' {\em US
  Patent 1,948,384} (1934) .

\bibitem{Lawrence:1936uw}
E.~O. Lawrence and D.~Cooksey, ``{On the apparatus for the multiple
  acceleration of light ions to high speed},''
  \href{http://dx.doi.org/10.1103/PhysRev.50.1131}{{\em Phys. Rev.} {\bfseries
  50} (1936) 1131--1140}.

\bibitem{Wideroe:1947egh}
R.~Wider{\"{o}}e, ``The `gigator'--a proposed new circular accelerator for
  heavy particles,'' {\em Phys. Rev.} {\bfseries 72} (1947) 978.

\bibitem{Wideroe:1953ufn}
R.~Wider{\"{o}}e, ``{Das Betatron},'' {\em Z. Angew. Phys.} {\bfseries 5}
  (1953) 187--200.

\bibitem{Neuffer:1994bt}
D.~V. Neuffer and R.~B. Palmer, ``{A High-Energy High-Luminosity $\mu^+ -
  \mu^-$ Collider},'' in {\em {Proceedings of the 4th European Particle
  Accelerator Conference}}.
\newblock 1994.

\bibitem{Geer:1997iz}
S.~Geer, ``{Neutrino beams from muon storage rings: Characteristics and physics
  potential},'' \href{http://dx.doi.org/10.1103/PhysRevD.57.6989}{{\em Phys.
  Rev.} {\bfseries D57} (1998) 6989--6997},
\href{http://arxiv.org/abs/hep-ph/9712290}{{\ttfamily arXiv:hep-ph/9712290}}.

\bibitem{Apollonio:2002en}
M.~Apollonio {\em et al.}, ``Oscillation physics with a neutrino factory,''
  \href{http://arXiv.org/abs/hep-ph/0210192}{{\ttfamily arXiv:hep-ph/0210192}}.

\bibitem{PhysRevSTAB.6.081001}
M.~M. Alsharo'a {\em et al.}, ``Recent progress in neutrino factory and muon
  collider research within the muon collaboration,''
  \href{http://dx.doi.org/10.1103/PhysRevSTAB.6.081001}{{\em Phys. Rev. ST
  Accel. Beams} {\bfseries 6} (2003) 081001}.

\bibitem{Palmer:2014nza}
R.~B. Palmer, ``{Muon Colliders},''
\href{http://dx.doi.org/10.1142/S1793626814300072}{{\em Rev. Accel. Sci. Tech.}
  {\bfseries 7} (2014) 137--159}.

\bibitem{Boscolo:2018tlu}
M.~Boscolo, M.~Antonelli, O.~R. Blanco-Garcia, S.~Guiducci, S.~Liuzzo,
  P.~Raimondi, and F.~Collamati, ``{Low emittance muon accelerator studies with
  production from positrons on target},''
  \href{http://dx.doi.org/10.1103/PhysRevAccelBeams.21.061005}{{\em Phys. Rev.
  Accel. Beams} {\bfseries 21} no.~6, (2018) 061005},
\href{http://arxiv.org/abs/1803.06696}{{\ttfamily arXiv:1803.06696
  [physics.acc-ph]}}.

\bibitem{Neuffer2018}
D.~Neuffer and V.~Shiltsev, ``On the feasibility of a pulsed 14 {TeV} c.m.e.
  muon collider in the {LHC} tunnel,''
  \href{http://dx.doi.org/10.1088/1748-0221/13/10/t10003}{{\em JINST}
  {\bfseries 13} no.~10, (2018) T10003--T10003}.

\bibitem{Behnke:2013xla}
T.~Behnke, J.~E. Brau, B.~Foster, J.~Fuster, M.~Harrison, J.~M. Paterson,
  M.~Peskin, M.~Stanitzki, N.~Walker, and H.~Yamamoto, ``{The International
  Linear Collider Technical Design Report - Volume 1: Executive Summary},''
\href{http://arxiv.org/abs/1306.6327}{{\ttfamily arXiv:1306.6327
  [physics.acc-ph]}}.

\bibitem{Charles:2018vfv}
{\bfseries CLIC and CLICdp} Collaboration, T.~K. Charles {\em et al.}, ``{The
  Compact Linear Collider (CLIC) - 2018 Summary Report},''
  \href{http://dx.doi.org/10.23731/CYRM-2018-002}{{\em CERN Yellow Rep.
  Monogr.} {\bfseries 1802} (2018) 1--98},
\href{http://arxiv.org/abs/1812.06018}{{\ttfamily arXiv:1812.06018
  [physics.acc-ph]}}.

\bibitem{Roloff:2018dqu}
{\bfseries CLIC and CLICdp} Collaboration, P.~Roloff, R.~Franceschini,
  U.~Schnoor, and A.~Wulzer, ``{The Compact Linear e$^+$e$^-$ Collider (CLIC):
  Physics Potential},''
\href{http://arxiv.org/abs/1812.07986}{{\ttfamily arXiv:1812.07986 [hep-ex]}}.

\bibitem{Aicheler:2019dhf}
{\bfseries CLIC accelerator} Collaboration, M.~Aicheler, P.~N. Burrows,
  N.~Catalan~Lasheras, R.~Corsini, M.~Draper, J.~Osborne, D.~Schulte,
  S.~Stapnes, and M.~J. Stuart, ``{The Compact Linear Collider (CLIC) - Project
  Implementation Plan},''
\href{http://arxiv.org/abs/1903.08655}{{\ttfamily arXiv:1903.08655
  [physics.acc-ph]}}.

\bibitem{Abada:2019zxq}
{\bfseries FCC} Collaboration, A.~Abada {\em et al.}, ``{FCC-ee: The Lepton
  Collider},''
\href{http://dx.doi.org/10.1140/epjst/e2019-900045-4}{{\em Eur. Phys. J.
  Special Topics} {\bfseries 228} (2019) 261--623}.

\bibitem{MYERS:2013hra}
S.~Myers, ``{The Large Hadron Collider 2008-2013},''
\href{http://dx.doi.org/10.1142/S0217751X13300354}{{\em Int. J. Mod. Phys.}
  {\bfseries A28} (2013) 1330035}.

\bibitem{2012acph.book.Lee}
S.~Y. {Lee}, \href{http://dx.doi.org/10.1142/8335}{{\em {Accelerator Physics
  (Third Edition)}}}.
\newblock World Scientific Publishing Co, 2012.

\bibitem{PhysRevLett.64.2901}
S.~Schr\"oder {\em et al.}, ``First laser cooling of relativistic ions in a
  storage ring,'' \href{http://dx.doi.org/10.1103/PhysRevLett.64.2901}{{\em
  Phys. Rev. Lett.} {\bfseries 64} (1990) 2901--2904}.

\bibitem{PhysRevLett.67.1238}
J.~S. Hangst, M.~Kristensen, J.~S. Nielsen, O.~Poulsen, J.~P. Schiffer, and
  P.~Shi, ``Laser cooling of a stored ion beam to 1 m{K},''
  \href{http://dx.doi.org/10.1103/PhysRevLett.67.1238}{{\em Phys. Rev. Lett.}
  {\bfseries 67} (1991) 1238--1241}.

\bibitem{doi:10.1063/1.329218}
P.~J. Channell, ``Laser cooling of heavy ion beams,''
  \href{http://dx.doi.org/10.1063/1.329218}{{\em Journal of Applied Physics}
  {\bfseries 52} no.~6, (1981) 3791--3793}.

\bibitem{Mohl:1980jb}
D.~Mohl, G.~Petrucci, L.~Thorndahl, and S.~Van Der~Meer, ``{Physics and
  Technique of Stochastic Cooling},''
\href{http://dx.doi.org/10.1016/0370-1573(80)90140-4}{{\em Phys. Rept.}
  {\bfseries 58} (1980) 73--119}.

\bibitem{Marriner:2003mn}
J.~Marriner, ``{Stochastic cooling overview},''
  \href{http://dx.doi.org/10.1016/j.nima.2004.06.025}{{\em Nucl. Instrum.
  Meth.} {\bfseries A532} (2004) 11--18},
\href{http://arxiv.org/abs/physics/0308044}{{\ttfamily arXiv:physics/0308044
  [physics]}}.

\bibitem{1063-7869-43-5-R01}
V.~V. Parkhomchuk and A.~N. Skrinsky, ``Electron cooling: 35 years of
  development,'' {\em Physics-Uspekhi} {\bfseries 43} no.~5, (2000) 433--452.
  \url{http://stacks.iop.org/1063-7869/43/i=5/a=R01}.

\bibitem{1999HyInt.119..305M}
M.~{M{\"u}hlbauer}, H.~{Daniel}, F.~J. {Hartmann}, P.~{Hauser}, F.~{Kottmann},
  C.~{Petitjean}, W.~{Schott}, D.~{Taqqu}, and P.~{Wojciechowski},
  ``{Frictional cooling: Experimental results},''
  \href{http://dx.doi.org/10.1023/A:1012624501134}{{\em Hyperfine Interactions}
  {\bfseries 119} (1999) 305--310}.

\bibitem{2005NIMPA.546..356A}
H.~{Abramowicz}, A.~{Caldwell}, R.~{Galea}, and S.~{Schlenstedt}, ``{A Muon
  Collider scheme based on Frictional Cooling},''
  \href{http://dx.doi.org/10.1016/j.nima.2005.03.125}{{\em Nucl. Instrum.
  Meth.} {\bfseries A546} (2005) 356--375},
  \href{http://arxiv.org/abs/physics/0410017}{{\ttfamily
  arXiv:physics/0410017}}.

\bibitem{2006PhRvL..97s4801T}
D.~{Taqqu}, ``{Compression and Extraction of Stopped Muons},''
  \href{http://dx.doi.org/10.1103/PhysRevLett.97.194801}{{\em Phys. Rev. Lett.}
  {\bfseries 97} no.~19, (2006) 194801}.

\bibitem{PhysRevLett.112.224801}
Y.~Bao, A.~Antognini, W.~Bertl, M.~Hildebrandt, K.~S. Khaw, K.~Kirch, A.~Papa,
  C.~Petitjean, F.~M. Piegsa, S.~Ritt, K.~Sedlak, A.~Stoykov, and D.~Taqqu,
  ``Muon cooling: Longitudinal compression,''
  \href{http://dx.doi.org/10.1103/PhysRevLett.112.224801}{{\em Phys. Rev.
  Lett.} {\bfseries 112} (2014) 224801}.

\bibitem{Skrinsky:1981zz}
A.~N. Skrinsky and V.~V. Parkhomchuk, ``{Cooling Methods for Beams of Charged
  Particles. (In Russian)},''
{\em Sov. J. Part. Nucl.} {\bfseries 12} (1981) 223--247.

\bibitem{Neuffer:1983jr}
D.~Neuffer, ``{Principles and Applications of Muon Cooling},''
{\em Part. Accel.} {\bfseries 14} (1983) 75--90.

\bibitem{Rogers:2013kna}
C.~T. Rogers, D.~Stratakis, G.~Prior, S.~Gilardoni, D.~Neuffer, P.~Snopok,
  A.~Alekou, and J.~Pasternak, ``{Muon front end for the neutrino factory},''
\href{http://dx.doi.org/10.1103/PhysRevSTAB.16.040104}{{\em Phys. Rev. ST
  Accel. Beams} {\bfseries 16} (2013) 040104}.

\bibitem{Stratakis:2014nna}
D.~Stratakis and R.~B. Palmer, ``{Rectilinear six-dimensional ionization
  cooling channel for a muon collider: A theoretical and numerical study},''
\href{http://dx.doi.org/10.1103/PhysRevSTAB.18.031003}{{\em Phys. Rev. ST
  Accel. Beams} {\bfseries 18} no.~3, (2015) 031003}.

\bibitem{Neuffer:2016gtw}
D.~Neuffer, H.~Sayed, J.~Acosta, D.~Summers, and T.~Hart, ``{Final Cooling for
  a High-Energy High-Luminosity Lepton Collider},''
  \href{http://dx.doi.org/10.1088/1748-0221/12/07/T07003}{{\em JINST}
  {\bfseries 12} no.~07, (2017) T07003},
\href{http://arxiv.org/abs/1612.08960}{{\ttfamily arXiv:1612.08960
  [physics.acc-ph]}}.

\bibitem{Bae:2018atj}
S.~Bae {\em et al.}, ``{First muon acceleration using a radio frequency
  accelerator},''
  \href{http://dx.doi.org/10.1103/PhysRevAccelBeams.21.050101}{{\em Phys. Rev.
  Accel. Beams} {\bfseries 21} no.~5, (2018) 050101},
\href{http://arxiv.org/abs/1803.07891}{{\ttfamily arXiv:1803.07891
  [physics.acc-ph]}}.

\bibitem{Mori:2009zz}
Y.~Mori, Y.~Ishi, Y.~Kuriyama, Y.~Sakurai, T.~Uesugi, K.~Okabe, and I.~Sakai,
  ``{Neutron Source with Emittance Recovery Internal Target},'' in {\em
  {Proceedings of the 23rd Particle Accelerator Conference}}.
\newblock 2009.
\newblock
  \url{http://accelconf.web.cern.ch/AccelConf/PAC2009/papers/th4gac04.pdf}.

\bibitem{MICE-WWW}
{\bfseries {MICE}} Collaboration, ``{International Muon Ionization Cooling
  Experiment}.'' \url{http://mice.iit.edu}.

\bibitem{Penn:2000mt}
G.~Penn and J.~S. Wurtele, ``Beam envelope equations for cooling of muons in
  solenoid fields,'' {\em Phys. Rev. Lett.} {\bfseries 85} (2000) 764.

\bibitem{holzer2004}
E.~B. Holzer, ``Figure of merit for muon cooling -- an algorithm for particle
  counting in coupled phase planes,''
  \href{http://dx.doi.org/10.1016/j.nima.2004.06.055}{{\em Nucl. Instrum.
  Meth.} {\bfseries A532} (2004) 270--274}.

\bibitem{Rogers08beamdynamics}
C.~Rogers, {\em Beam Dynamics in an Ionisation Cooling Channel}.
\newblock {PhD} dissertation, Imperial College, London, 2008.

\bibitem{Mack19791}
Y.~Mack and M.~Rosenblatt, ``Multivariate k-nearest neighbor density
  estimates,''
  \href{http://dx.doi.org/https://doi.org/10.1016/0047-259X(79)90065-4}{{\em
  Journal of Multivariate Analysis} {\bfseries 9} no.~1, (1979) 1 -- 15}.

\bibitem{drielsma:2018}
F.~Drielsma,
  \href{http://dx.doi.org/10.13097/archive-ouverte/unige:114100}{{\em
  Measurement of the increase in phase space density of a muon beam through
  ionization cooling}}.
\newblock PhD thesis, University of Geneva, 2018.

\bibitem{Bogomilov:2012sr}
{\bfseries MICE} Collaboration, M.~Bogomilov {\em et al.}, ``{The MICE Muon
  Beam on ISIS and the beam-line instrumentation of the Muon Ionization Cooling
  Experiment},'' \href{http://dx.doi.org/10.1088/1748-0221/7/05/P05009}{{\em
  JINST} {\bfseries 7} (2012) P05009},
\href{http://arxiv.org/abs/1203.4089}{{\ttfamily arXiv:1203.4089
  [physics.acc-ph]}}.

\bibitem{Adams:2013lba}
{\bfseries MICE} Collaboration, D.~Adams {\em et al.}, ``{Characterisation of
  the muon beams for the Muon Ionisation Cooling Experiment},''
  \href{http://dx.doi.org/10.1140/epjc/s10052-013-2582-8}{{\em Eur. Phys. J.}
  {\bfseries C73} no.~10, (2013) 2582},
\href{http://arxiv.org/abs/1306.1509}{{\ttfamily arXiv:1306.1509
  [physics.acc-ph]}}.

\bibitem{Adams:2015wxp}
{\bfseries MICE} Collaboration, M.~Bogomilov {\em et al.}, ``{Pion
  Contamination in the MICE Muon Beam},''
  \href{http://dx.doi.org/10.1088/1748-0221/11/03/P03001}{{\em JINST}
  {\bfseries 11} no.~03, (2016) P03001},
\href{http://arxiv.org/abs/1511.00556}{{\ttfamily arXiv:1511.00556
  [physics.ins-det]}}.

\bibitem{Booth_2016}
C.~Booth, P.~Hodgson, J.~Langlands, E.~Overton, M.~Robinson, P.~Smith,
  G.~Barber, K.~Long, B.~Shepherd, E.~Capocci, C.~MacWaters, and J.~Tarrant,
  ``The design and performance of an improved target for {MICE},''
  \href{http://dx.doi.org/10.1088/1748-0221/11/05/p05006}{{\em JINST}
  {\bfseries 11} no.~05, (2016) P05006--P05006}.

\bibitem{THOMASON201961}
J.~Thomason, ``{The ISIS Spallation Neutron and Muon Source - The first
  thirty-three years},''
  \href{http://dx.doi.org/https://doi.org/10.1016/j.nima.2018.11.129}{{\em
  Nucl. Instrum. Meth.} {\bfseries A917} (2019) 61 -- 67}.

\bibitem{Bayliss:2018rkd}
{\bfseries MICE} Collaboration, V.~Bayliss {\em et al.}, ``{The liquid-hydrogen
  absorber for MICE},''
  \href{http://dx.doi.org/10.1088/1748-0221/13/09/T09008}{{\em JINST}
  {\bfseries 13} no.~09, (2018) T09008},
\href{http://arxiv.org/abs/1807.03019}{{\ttfamily arXiv:1807.03019
  [physics.acc-ph]}}.

\bibitem{PhysRevSTAB.18.044201}
D.~Stratakis, R.~B. Palmer, and D.~P. Grote, ``Influence of space-charge fields
  on the cooling process of muon beams,''
  \href{http://dx.doi.org/10.1103/PhysRevSTAB.18.044201}{{\em Phys. Rev. ST
  Accel. Beams} {\bfseries 18} (2015) 044201}.

\bibitem{Adams2019}
{\bfseries MICE} Collaboration, V.~Blackmore {\em et al.}, ``First
  particle-by-particle measurement of emittance in the muon ionization cooling
  experiment,'' \href{http://dx.doi.org/10.1140/epjc/s10052-019-6674-y}{{\em
  Eur. Phys. J. C} {\bfseries 79} no.~3, (2019) 257}.

\bibitem{Bertoni:2010by}
{\bfseries MICE} Collaboration, R.~Bertoni {\em et al.}, ``{The design and
  commissioning of the MICE upstream time-of-flight system},''
  \href{http://dx.doi.org/10.1016/j.nima.2009.12.065}{{\em Nucl. Instrum.
  Meth.} {\bfseries A615} (2010) 14--26},
\href{http://arxiv.org/abs/1001.4426}{{\ttfamily arXiv:1001.4426
  [physics.ins-det]}}.

\bibitem{Bertoni:2010bb}
R.~Bertoni, M.~Bonesini, A.~deBari, G.~Cecchet, Y.~Karadzhov, and R.~Mazza,
  ``{The construction of the MICE TOF2 detector},'' {\em MICE Technical Note
  254} (2010) .
  \url{http://mice.iit.edu/micenotes/public/pdf/MICE0286/MICE0286.pdf}.

\bibitem{Cremaldi:2009}
L.~Cremaldi, D.~Sanders, P.~Sonnek, D.~Summers, and J.~Reidy, ``A cherenkov
  radiation detector with high density aerogels,''
  \href{http://dx.doi.org/10.1109/TNS.2009.2021266}{{\em IEEE Transactions on
  Nuclear Science} {\bfseries 56} (2009) 1475 -- 1478}.

\bibitem{Ellis:2010bb}
M.~Ellis {\em et al.}, ``{The Design, construction and performance of the MICE
  scintillating fibre trackers},''
  \href{http://dx.doi.org/10.1016/j.nima.2011.04.041}{{\em Nucl. Instrum.
  Meth.} {\bfseries A659} (2011) 136--153},
\href{http://arxiv.org/abs/1005.3491}{{\ttfamily arXiv:1005.3491
  [physics.ins-det]}}.

\bibitem{Dobbs:2016ejn}
A.~Dobbs, C.~Hunt, K.~Long, E.~Santos, M.~A. Uchida, P.~Kyberd, C.~Heidt,
  S.~Blot, and E.~Overton, ``{The reconstruction software for the MICE
  scintillating fibre trackers},''
  \href{http://dx.doi.org/10.1088/1748-0221/11/12/T12001}{{\em JINST}
  {\bfseries 11} no.~12, (2016) T12001},
\href{http://arxiv.org/abs/1610.05161}{{\ttfamily arXiv:1610.05161
  [physics.ins-det]}}.

\bibitem{Adams:2015eva}
{\bfseries MICE} Collaboration, D.~Adams {\em et al.}, ``{Electron-Muon Ranger:
  performance in the MICE Muon Beam},''
  \href{http://dx.doi.org/10.1088/1748-0221/10/12/P12012}{{\em JINST}
  {\bfseries 10} no.~12, (2015) P12012},
\href{http://arxiv.org/abs/1510.08306}{{\ttfamily arXiv:1510.08306
  [physics.ins-det]}}.

\bibitem{Asfandiyarov:2016erh}
R.~Asfandiyarov {\em et al.}, ``{The design and construction of the MICE
  Electron-Muon Ranger},''
  \href{http://dx.doi.org/10.1088/1748-0221/11/10/T10007}{{\em JINST}
  {\bfseries 11} no.~10, (2016) T10007},
\href{http://arxiv.org/abs/1607.04955}{{\ttfamily arXiv:1607.04955
  [physics.ins-det]}}.

\bibitem{VLPC}
M.~Petroff and M.~Stapelbroek, ``{Photon-Counting Solid-State
  Photomultiplier},'' {\em {IEEE Transactions on Nuclear Science}} {\bfseries
  {36}} no.~{1, Part 1}, ({1989}) {158--162}.

\bibitem{VLPC1}
M.~Petroff and M.~Atac, ``{High-Energy Particle Tracking using Scintillation
  Fibers and Solid-State Photomultipliers},'' {\em {IEEE Transactions on
  Nuclear Science}} {\bfseries {36}} no.~{1, Part 1}, ({1989}) {163--164}.

\bibitem{Agostinelli:2002hh}
S.~Agostinelli {\em et al.}, ``{GEANT4: A Simulation toolkit},''
\href{http://dx.doi.org/10.1016/S0168-9002(03)01368-8}{{\em Nucl. Instrum.
  Meth.} {\bfseries A506} (2003) 250--303}.

\bibitem{Alli06}
J.~Allison {\em et al.}, ``Geant4 developments and applications,''
  \href{http://dx.doi.org/10.1109/TNS.2006.869826}{{\em IEEE Transactions on
  Nuclear Science} {\bfseries 53} (2006) 270}.

\bibitem{Asfandiyarov:2018yds}
R.~Asfandiyarov {\em et al.}, ``{MAUS: The MICE Analysis User Software},'' {\em
  JINST} {\bfseries 14} (2019) T04005--T04005,
  \href{http://arxiv.org/abs/1812.02674}{{\ttfamily arXiv:1812.02674
  [physics.comp-ph]}}.

\end{thebibliography}\endgroup

\clearpage

\begin{multicols}{2}
  \appendix
  \section*{Methods}

\noindent\textbf{Data-taking and reconstruction} \\
\noindent
Data were buffered in the front-end electronics and read out after each 
target actuation. Data storage was triggered by a coincidence of signals
in the photmultiplier tubes (PMTs) serving a single scintillator slab in TOF1. 
The data recorded in response to a particular trigger are referred to
as a `particle event'.

Each TOF station was composed of a number of scintillator slabs that 
were read out using a pair of PMTs, one mounted at each end of the 
slab.
The reconstruction of the data began with the search for coincidences
in the signals from the two PMTs serving each slab in each TOF plane.
Such coincidences were referred to as `slab hits'.
`Space points' were then formed from the intersection of slab hits in
the $x$ and $y$ projections of each TOF station separately.
The position and time at which a particle giving rise to the space
point crossed the TOF station was then calculated using the slab
position and the times measured in each of the PMTs.
The relative timing of TOF0 and TOF1 was calibrated relative to the
observed time taken for electrons to pass between the two detectors,
on the assumption that they travelled at the speed of light.

Signals in the tracker readout were collected to reconstruct the
helical trajectories (`tracks') of charged particles in the upstream
and downstream trackers (TKU and TKD respectively).
Multiple Coulomb scattering introduced significant uncertainties in the
reconstruction of the helical trajectory of tracks with a bending
radius less than 5\,mm.
For this class of track momentum was deduced by combining the tracker
measurement with the measurements from nearby detectors.
Track-fit quality was characterised by the $\chi^2$ per degree-of-freedom

\begin{equation}
\chi^2_{\rm df} = \frac{1}{n} \sum_i \frac{\delta x^2_i}{\sigma^2_i}\,
\end{equation}

\noindent 
where $\delta x_i$ is the distance between the fitted track and the
measured signal in the $i^{th}$ tracker plane, $\sigma_i$ is the
resolution of the position measurement in the tracker planes and $n$
is the number of planes that had a signal used in the track
reconstruction. Further details of the reconstruction and simulation may be found in~\cite{Asfandiyarov:2018yds}. \\

\noindent\textbf{Beam selection} \\
\noindent
Measurements made in the instrumentation upstream of the absorber were
used to select the input beam for the study of ionization cooling
presented in this paper.
The input beam (the `upstream sample') was composed of those events
that satisfied the following criteria:
\begin{itemize}
  \item Exactly one space point was found in TOF0 and TOF1 and exactly one track
    in TKU;
  \item The track in TKU had $\chi^2_{\rm df} < 8$ and was contained within
   the 150\,mm fiducial radius over the full length of the tracker;
  \item The track in TKU had a reconstructed momentum in the
   range 135--145\,MeV/c corresponding to the momentum acceptance of
   the cooling cell;
   \item The time-of-flight between TOF0 and TOF1 was consistent with
    that of a muon given the momentum measured in TKU; and
  \item The radius at which the track in TKU passed through the
    diffuser was smaller than the diffuser aperture.
\end{itemize}
The beam emerging from the cooling cell (the `downstream sample') was
characterised using the subset of the upstream sample that satisfied
the following criteria:
\begin{itemize}
  \item Exactly one track was found in TKD; and
  \item The track in TKD had a $\chi^2_{\rm df} < 8$ and was
    contained within the 150\,mm fiducial radius of TKD over the full
    length of the tracker. 
\end{itemize}
The same sample-selection criteria were used to select events from the
simulation of the experiment, which includes a reconstruction of the
electronics signals expected for the simulated particles. \\

\noindent\textbf{Correction for detector effects} \\
\noindent
The amplitude distributions obtained from the upstream and downstream
samples were corrected for the effects of detector efficiency and
resolution and to take account of migration of events between
amplitude bins.
The corrected number of events in a bin, $N_i^{corr}$, was calculated
from the raw number of events, $N_{j}^{raw}$, using

\begin{equation}
  N_i^{corr} = E_{i} \sum_j S_{ij} N_{j}^{raw},
\end{equation}

\noindent 
where $E_{i}$ is the efficiency-correction factor and $S_{ij}$
accounts for detector resolution and event migration.
$E_{i}$ and $S_{ij}$ were estimated from the simulation of the
experiment.
The uncorrected and corrected amplitude distributions for a particular configuration 
are shown in figure~\ref{fig:amplitude_pdf_reco_correction}. 
The correction is small relative to the ionization cooling effect; the
ionization cooling effect is clear even in the uncorrected distributions.

\end{multicols}
\setcounter{figure}{0}
\begin{figure*}[h]
  \begin{center}
    \includegraphics[width=0.75\textwidth]{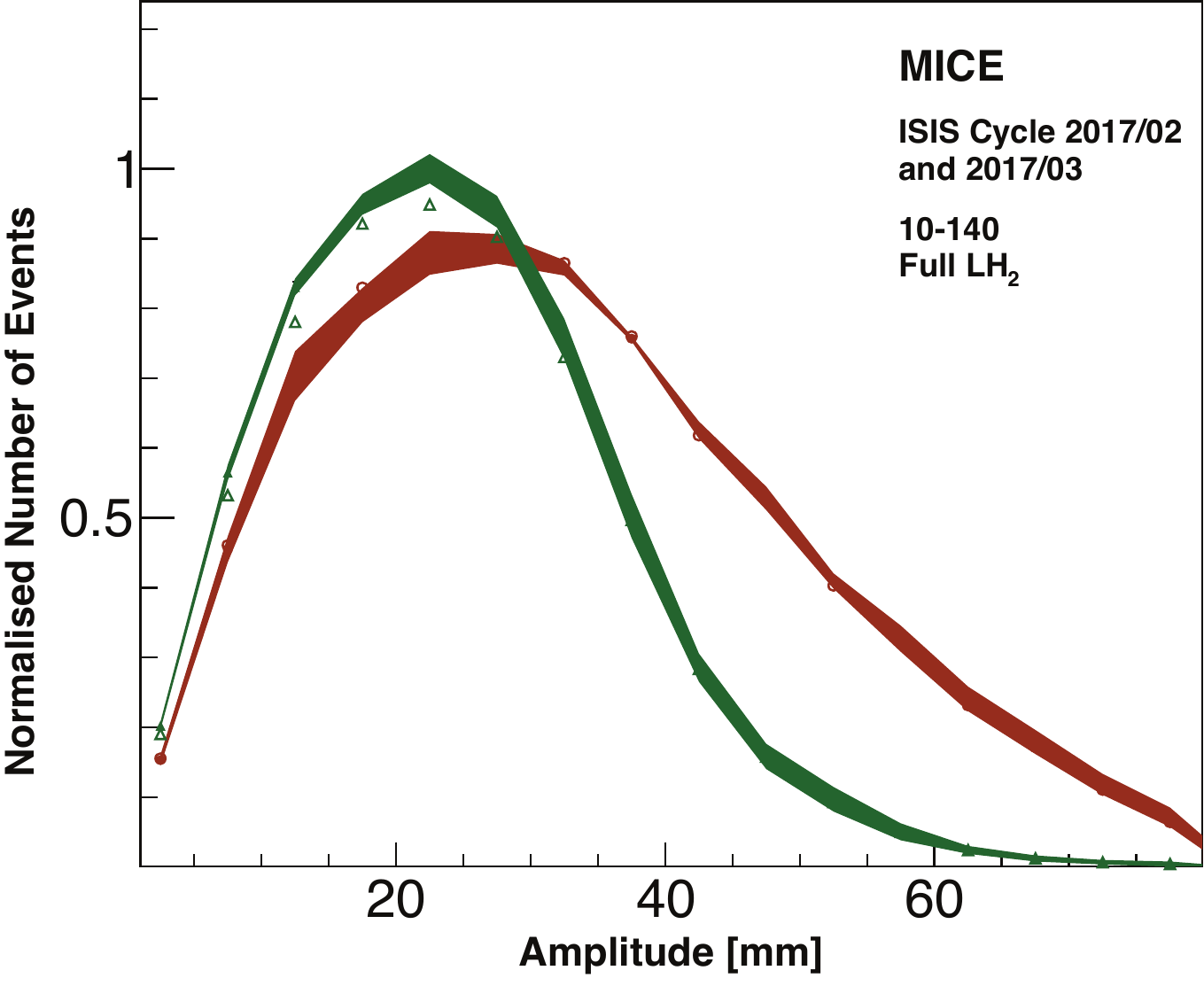}
  \end{center}
  \caption{
    Distribution of amplitudes with corrected and uncorrected distribution shown for the 10-140 LH$_2$ full configuration. 
    The uncorrected data is shown by open points while the corrected data is shown
    by filled points. The upstream distribution is shown by orange circles while the
    downstream distribution is shown by green triangles. 
    Systematic uncertainty is shown by coloured bands. Statistical error is shown by bars and is
    just visible for a few points. 
  }
  \label{fig:amplitude_pdf_reco_correction}
\end{figure*}

\clearpage
\appendix
%
\thispagestyle{plain}
\setlength\parindent{0em}

\section*{The MICE collaboration}

M.~Bogomilov,  R.~Tsenov, G.~Vankova-Kirilova
\\{\it
   Department of Atomic Physics, St.~Kliment Ohridski University of Sofia, Sofia, Bulgaria
}\\

Y.~P.~Song, J.~Y.~Tang
\\{\it
Institute of High Energy Physics, Chinese Academy of Sciences, Beijing, China
}\\

Z.~H.~Li
\\{\it
Sichuan University, China
}\\

R.~Bertoni, M.~Bonesini, F.~Chignoli, R.~Mazza
\\{\it
Sezione INFN Milano Bicocca, Dipartimento di Fisica G.~Occhialini, Milano, Italy
}\\

V.~Palladino
\\{\it
Sezione INFN Napoli and Dipartimento di Fisica, Universit\`{a} Federico II, Complesso Universitario di Monte S.~Angelo, Napoli, Italy
}\\

A.~de Bari
\\{\it 
Sezione INFN Pavia and Dipartimento di Fisica, Pavia, Italy
}\\

D.~Orestano, L.~Tortora
\\{\it
INFN Sezione di Roma Tre and Dipartimento di Matematica e Fisica, Universit\`{a} Roma Tre, Italy
}\\

Y.~Kuno, H.~Sakamoto\footnote{Current address RIKEN 2-1 Horosawa, Wako, Saitama 351-0198, Japan}, A.~Sato
\\{\it
Osaka University, Graduate School of Science, Department of Physics, Toyonaka, Osaka, Japan
}\\

S.~Ishimoto
\\{\it
High Energy Accelerator Research Organization (KEK), Institute of Particle and Nuclear Studies, Tsukuba, Ibaraki, Japan
}\\

M.~Chung, C.~K.~Sung
\\{\it 
UNIST, Ulsan, Korea
}\\

F.~Filthaut\footnote{Also at Radboud University, Nijmegen, The Netherlands}
\\{\it
Nikhef, Amsterdam, The Netherlands
}\\

D.~Jokovic, D.~Maletic, M.~Savic
\\{\it
Institute of Physics, University of Belgrade, Serbia
}\\

N.~Jovancevic, J.~Nikolov
\\{\it
Faculty of Sciences,  University of Novi Sad,  Serbia
}\\

\newpage
M. ~Vretenar, S.~Ramberger
\\{\it
  CERN, Esplanade des Particules 1, P.O. Box 1211, Geneva 23, Switzerland
}\\

R.~Asfandiyarov, A.~Blondel, F.~Drielsma, Y.~Karadzhov 
\\{\it
DPNC, Section de Physique, Universit\'e de Gen\`eve, Geneva, Switzerland
}\\

G.~Charnley, N.~Collomb,  K.~Dumbell, A.~Gallagher, A.~Grant, S.~Griffiths,  T.~Hartnett, B.~Martlew, 
A.~Moss, A.~Muir, I.~Mullacrane, A.~Oates, P.~Owens, G.~Stokes, P.~Warburton, C.~White
\\{\it
STFC Daresbury Laboratory, Daresbury, Cheshire, UK
}\\

D.~Adams,   V.~Bayliss, J.~Boehm, T.~W.~Bradshaw, C.~Brown\footnote{Also at Brunel University, Uxbridge, UB8 3PH, UK}, M.~Courthold,  J.~Govans, M.~Hills, J.-B.~Lagrange, C.~Macwaters, A.~Nichols, R.~Preece, S.~Ricciardi, C.~Rogers, T.~Stanley, J.~Tarrant,  
M.~Tucker, S.~Watson\footnote{Current address ATC, Royal Observatory Edinburgh, Blackford Hill,  Edinburgh EH9 3HJ}, A.~Wilson
\\{\it
 STFC Rutherford Appleton Laboratory, Harwell Oxford, Didcot, UK
}\\

R.~Bayes\footnote{Current address Laurentian University, 935 Ramsey Lake Road, Sudbury, ON, Canada},  J.~C.~Nugent, F.~J.~P.~Soler
\\{\it
School of Physics and Astronomy, Kelvin Building, The University of Glasgow, Glasgow, UK
}\\

R.~Gamet, P.~Cooke
\\{\it
Department of Physics, University of Liverpool, Liverpool, UK
}\\

V.~J.~Blackmore, D.~Colling, A.~Dobbs\footnote{Current address OPERA Simulation Software, Network House, Langford Locks, Kidlington, Oxfordshire, OX5 1LH, UK}, P.~Dornan, P.~Franchini, C.~Hunt\footnote{Current address CERN, Esplanade des Particules 1, P.O. Box, 1211 Geneva 23, Switzerland.}, P.~B.~Jurj, A.~Kurup, K.~Long, J.~Martyniak,  S.~Middleton\footnote{Current address School of Physics and Astronomy, University of Manchester, Oxford Road, Manchester M13 9PL, UK}, J.~Pasternak, M.~A.~Uchida\footnote{Current address Rutherford Building, Cavendish Laboratory, JJ Thomson Avenue, Cambridge CB3 0HE, UK}
\\{\it
Department of Physics, Blackett Laboratory, Imperial College London, London, UK
}\\

J.~H.~Cobb
\\{\it
Department of Physics, University of Oxford, Denys Wilkinson Building, Oxford, UK
}\\

C.~N.~Booth, P.~Hodgson, J.~Langlands, E.~Overton\footnote{Current address Arm, City Gate, 8 St Mary's Gate, Sheffield, S1 4LW, United Kingdom}, V.~Pec,  P.~J.~Smith, S.~Wilbur
\\{\it
Department of Physics and Astronomy, University of Sheffield, Sheffield, UK
}\\

G.~T.~Chatzitheodoridis\footnote{Also at School of Physics and Astronomy, Kelvin Building, The University of Glasgow, Glasgow, UK}$^,$\footnotemark, A.~J.~Dick\footnotemark[\value{footnote}],  K.~Ronald\footnotemark[\value{footnote}], C.~G.~Whyte\footnotemark[\value{footnote}], A.~R.~Young\footnotemark[\value{footnote}]
\footnotetext{Also at Cockcroft Institute, Daresbury Laboratory, Sci-Tech Daresbury, Daresbury, Warrington, WA4 4AD, UK}
\\{\it
SUPA and the Department of Physics, University of Strathclyde, Glasgow, UK
}\\

S.~Boyd,  J.~R.~Greis, T.~Lord, C.~Pidcott\footnote{Current address Department of Physics and Astronomy, University of Sheffield, Sheffield, UK}, I.~Taylor\footnote{Current address Defence Science and Technology Laboratory, Salisbury, SP4 0JQ, UK}
\\{\it
Department of Physics, University of Warwick, Coventry, UK
}\\

M.~Ellis\footnote{Current address Westpac Group, Sydney, Australia}, R.B.S.~Gardener, P.~Kyberd, J.~J.~Nebrensky
\\{\it
Brunel University, Uxbridge, UB8 3PH, UK
}\\

M.~Palmer, H.~Witte
\\{\it
Brookhaven National Laboratory, NY, USA
}\\

\newcounter{FNEuclid}
D.~Adey\footnote{Current address Institute of High Energy Physics, Chinese Academy of Sciences, Bejing, China}, A.~D.~Bross, D.~Bowring, P.~Hanlet, A.~Liu\footnotemark\footnotetext{Current address Euclid Techlabs, Bolingbrook, Illinois 60440, USA}\setcounter{FNEuclid}{\value{footnote}}, D.~Neuffer, M.~Popovic, P.~Rubinov
\\{\it
Fermilab, PO Box 500, Batavia IL 60510-5011, USA
}\\

A.~DeMello, S.~Gourlay, A.~Lambert, D.~Li, T.~Luo, S.~Prestemon,  S.~Virostek
\\{\it
Lawrence Berkeley National Laboratory, Berkeley, CA, USA
}\\

B.~Freemire\footnotemark[\value{FNEuclid}], D.~M.~Kaplan, T.~A.~Mohayai\footnote{Current address Fermilab, PO Box 500, Batavia IL 60510-5011, USA}, D.~Rajaram\footnote{Current address Illinois Institute of Technology, College of Science, Robert A. Pritzker Science Center, 3105 South Dearborn, Chicago, IL 60616P, USA}, P.~Snopok, Y.~Torun
\\{\it
Illinois Institute of Technology, Chicago, IL, USA
}\\

L.~M.~Cremaldi, D.~A.~Sanders, D.~J.~Summers
\\{\it
University of Mississippi, Oxford, MS, USA
}\\

L.~R.~Coney\footnote{Current address European Spallation Source ERIC, Box 176, SE-221 00 Lund, Sweden}, G.~G.~Hanson, C.~Heidt.
\\{\it
University of California, Riverside, CA, USA
}\\

\end{document}